# THE HYDROGEN ATOM ACCORDING TO RELATIVISTIC WAVE MECHANICS -- AMPLITUDE FUNCTIONS AND CIRCULATING ELECTRONIC CURRENTS

J. F. Ogilvie*

Centre for Experimental and Constructive Mathematics, Department of Mathematics,
Simon Fraser University, Burnaby, British Columbia V5A 1S6, Canada
Escuela de Química, Universidad de Costa Rica, Ciudad Universitaria Rodrigo Facio,
San Pedro de Montes de Oca, San José 11501-2050, Costa Rica

**Abstract**

The solution of Dirac's equation for the hydrogen atom according to relativistic wave mechanics yields for each state a vectorial amplitude function with four components, two large and two small. Each such component has its characteristic surface of constant amplitude, of which we plot several examples. For each state of the hydrogen atom there is both a density of electronic charge surrounding the atomic nucleus and an electronic current circulating about the polar axis; the latter generates a magnetic dipolar moment that agrees precisely with experiment.

## I  INTRODUCTION

By the time that Schroedinger developed his wave-mechanical method of quantum mechanics with a calculation of the hydrogen atom in discrete states that derived energies proportional to an inverse square of integers [1], measurements of atomic spectra had already proved that that result was inadequate: cooling hydrogen atoms near the temperature of liquid helium, Shrum confirmed that the spectral lines in the Balmer series were not singlets but multiplets [2]. Within two years of Schroedinger's publication [1], Dirac formulated his relativistic wave mechanics [3], by means of which Darwin [4] and Gordon [5] solved the Dirac equation to produce energies of the hydrogen atom that explained these multiplets.

The important advance of Dirac's equation beyond that of Schroedinger was that spatial coordinates and time were treated in an equitable manner, as required according to the theory of special relativity. Whereas Schroedinger's temporally dependent equation contained second derivatives of the three spatial coordinates but a first derivative with respect to time [1], Dirac's corresponding equation encompassed first derivatives of all four quantities [3]; to satisfy the required properties, the resulting amplitude function could not be a scalar quantity but had to be a vectorial quantity with four components. The density of electronic charge, which is proportional to the square of Schroedinger's scalar amplitude function for the hydrogen atom, or precisely a product of a function and its complex conjugate as $\psi^* \psi$ to take account of the complex character, hence became proportional to a scalar product of Dirac's vectorial amplitude function for each state with its complex conjugate vector.

Our objective in this work is to illustrate some important aspects of amplitude functions of the hydrogen atom according to Dirac's theory, as described by Powell [6]. We illustrate selected amplitude functions and their squares, for comparison with the amplitude functions according to Schroedinger's non-relativistic wave mechanics in four coordinate systems [7 - 11], and discuss the probability flux for the Dirac functions in relation to the magnetic properties of the hydrogen atom. Here we present only the results of the essential calculations; the details of these calculations are available elsewhere [12].

## II AMPLITUDE FUNCTIONS

A significant advantage of the use of Dirac's amplitude functions for the hydrogen atom is that there is an unique identification of each state of the hydrogen atom, characterised with its energy and angular momentum, with a particular amplitude function, characterized in turn with four quantum numbers -- $n$, $l$, $j$, $m_j$ according to the following meanings. Quantum number $n$ is the principal quantum number, closely related to energy quantum number $n$ [7] that arose from the interpretation, by Balmer and Rydberg, of experimental measurements of spectral lines emitted from hydrogen discharges; the only acceptable values of $n$ are positive integers. Although $l$ in Schroedinger's wave mechanics for the hydrogen atom in spherical polar coordinates is related to angular momentum in that the square of that quantity is proportional to $l(l + 1)$, in Dirac's context in the same spherical polar coordinates, $l$ is simply an azimuthal quantum number, which assumes values of non-negative integers. Angular-momentum quantum number $j$ assumes solely positive values numbering at most two, only $|l \pm ½|$, so half the value of an odd positive integer. Magnetic quantum number $m_j$ assumes all values from $-j$ to $+j$, so half the value of an odd negative or odd positive integer. We apply these four quantum numbers within a *ket*, printed as $|n,l,j,m_j>$ according to the notation that Dirac devised, to specify both a state of the hydrogen atom and its associated amplitude function.

For the electronic ground states of the hydrogen atom, so $|1,0,½,½>$ or $|1,0,½,-½>$, which are degenerate in the absence of an externally applied magnetic field, we investigate the properties of this amplitude function.

$$|1,0,\tfrac{1}{2},\tfrac{1}{2}\rangle \;=\; \begin{bmatrix} e^{(-r)} \\ 0 \\ \dfrac{1}{2} i\,\alpha\, Z\, e^{(-r)} \cos(\theta) \\ \dfrac{1}{2} i\,\alpha\, Z\, e^{(-r)} \sin(\theta)\, e^{(i\phi)} \end{bmatrix}$$

Therein for the four components of the vector displayed in column form, apart from symbols $r,\theta,\phi$ for spherical polar coordinates with their conventional meanings, appear $\mathbf{i} = \sqrt{-1}$ as the imaginary unit, atomic number $Z$ with $Z=1$ for hydrogen, and fine-structure constant $\alpha$, of value $\sim 1/137$. For this amplitude function the second component is zero, but the first and second components are invariably distinguished from the third and fourth components that, when not zero, contain $\alpha$; for this reason the upper two components are called large components and the other two components are called small components. This distinction is valid only when $Z$ is small; for $Z \sim 100$, the magnitudes of all four components, if not zero, become comparable. Each amplitude function presented here is not normalised and represents only the leading term in a series [12]; as further terms contain factors $\alpha Z$ to even powers, their contributions to the total function are negligible for $Z$ small. We take $Z=1$ in all calculations and for all plots.

When we examine the four components of $|1,0,\tfrac{1}{2},\tfrac{1}{2}\rangle$, we perceive that the second component is zero and the first component contains no angular coordinate; both the third and fourth components contain $\mathbf{i}$ so that the third component is purely imaginary and the fourth component has both real and imaginary parts through the presence of $\mathbf{e}^{i\phi}$. We plot these components according to the same scheme applied elsewhere [7 - 11]: we calculate a surface corresponding to the formula with a fixed value of amplitude; this fixed value is taken as 1/100 of the maximum value of that function, which, for a single function, would imply 0.995 of the total electronic charge density to be enclosed within that surface. The scale along each cartesian axis has unit Bohr radius $a_0$ in all succeeding plots of these surfaces. For illustrative purposes, the plots of all small components and their squares are scaled by $(\alpha Z)^{-1}$, so that their sizes appear comparable with those of the large components. Figure 1a shows a plot of the first component of $|1,0,\tfrac{1}{2},\tfrac{1}{2}\rangle$; figure 1b shows the imaginary part of the third component, i.e. the coefficient of i, and figures 1c and 1d show the real and imaginary parts of the fourth component, respectively; the fourth component contains $\mathbf{e}^{i\phi}$ that becomes $\cos(\phi) + \mathbf{i} \sin(\phi)$ with its real and imaginary parts. In each case in which comparison is practicable, we relate these surfaces with surfaces of non-relativistic amplitude functions $\psi_{k,l,m}(r,\theta,\phi)$ plotted according to the same criteria, such as those presented in a collection elsewhere [8].

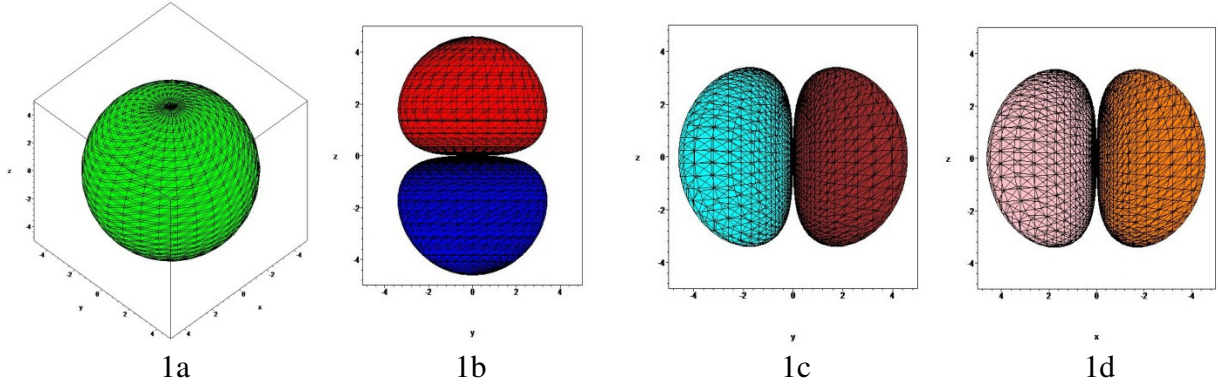

1a        1b        1c        1d

Figure 1. Surfaces of components of amplitude function $|1,0,½,½\rangle$; (a) first component, real; (b) third component, imaginary; (c) real part of fourth component; (d) imaginary part of fourth component.

The spherical shape of the surface of the first and large component reflects its lack of angular dependence, which clearly resembles the surface of $\psi_{0,0,0}(r,\theta,\phi)$. For the small components, the imaginary third component with its two nearly hemispherical lobes cylindrically symmetric about axis $z$ resembles the surface of $\psi_{0,1,0}(r,\theta,\phi)$, which is entirely real, whereas the surfaces of the real and imaginary parts of the fourth component resemble the corresponding parts of $\psi_{0,1,1}(r,\theta,\phi)$ with their nearly hemispherical lobes about axes $y$ and $x$, respectively. The square of this amplitude function is formed as a scalar product of the vector and its complex conjugate,

$$|1,0,½,½\rangle^* \cdot |1,0,½,½\rangle = \frac{1}{4} e^{(-2r)} (Z^2 \alpha^2 + 4),$$

which clearly has its own large and small parts, the latter containing $\alpha^2$, but is entirely real; the latter property is essential because the density of electronic charge, which must be a real quantity, is proportional to this scalar product. Both terms of this sum plot as spheres, but the part containing $\alpha^2$ makes a minuscule contribution to the total volume of the complete square of the amplitude function: on the same scale and for $Z=1$, the sphere of the small part would be smaller than the proton as atomic nucleus. Because these functions are not normalised, the resulting sphere of the total function would appear to be smaller than the sphere in figure 1a, but this feature is of no consequence for a qualitative understanding of the properties.

The amplitude function for the ground state with $m_j = -½$ has this form,

$$|1,0,½,-½\rangle = \begin{bmatrix} 0 \\ -e^{(-r)} \\ -\frac{1}{2} i \alpha Z e^{(-r)} \sin(\theta) e^{(-i\phi)} \\ \frac{1}{2} i \alpha Z e^{(-r)} \cos(\theta) \end{bmatrix},$$

which, on comparison with the components of $|1,0,½,½\rangle$ above, displays reversed signs, which amount to a change of phase from positive to negative or vice versa, and the reversal of the order within the two large or small components. With these reversals taken into consideration, the surfaces of the components of $|1,0,½,-½\rangle$ and of its square are obviously analogous to those of $|1,0,½,½\rangle$.

Amplitude function $|2,0,½,½\rangle$ has this form.

$$|2,0,½,½\rangle = \begin{bmatrix} \frac{1}{4}\sqrt{2}\left(1-\frac{r}{2}\right)e^{\left(-\frac{r}{2}\right)} \\ 0 \\ \frac{1}{8}i\sqrt{2}\,\alpha\,Z\left(1-\frac{r}{4}\right)e^{\left(-\frac{r}{2}\right)}\cos(\theta) \\ \frac{1}{8}i\sqrt{2}\,\alpha\,Z\left(1-\frac{r}{4}\right)e^{\left(-\frac{r}{2}\right)}\sin(\theta)\,e^{(i\phi)} \end{bmatrix}$$

Like $|1,0,½,½\rangle$, this amplitude function has a zero second component and a real first component; the third and small component is purely imaginary and the fourth component, also small, has both real and imaginary parts. Figure 2a shows a plot of the first component; figure 2b shows the imaginary part of the third component, i.e. the coefficient of **i**, and figures 2c and 2 d show the real and imaginary parts of the fourth component, respectively.

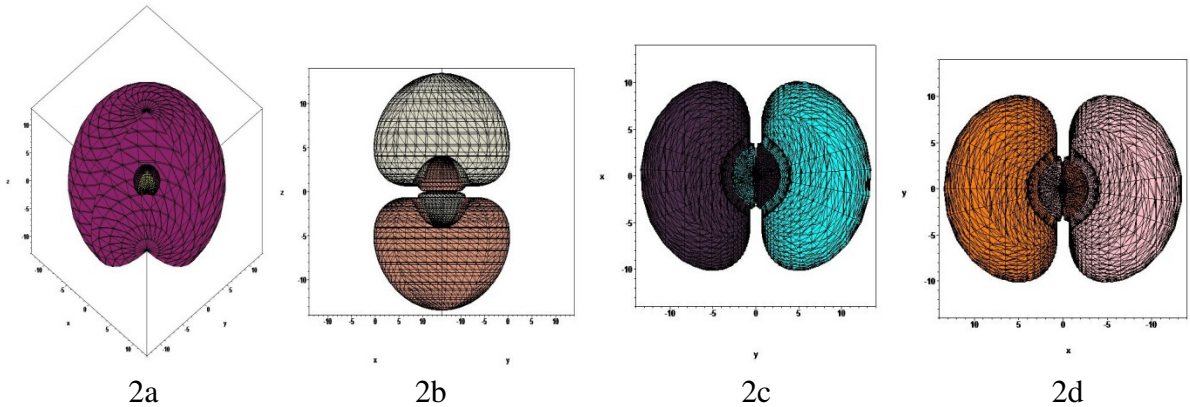

2a         2b         2c         2d

Figure 2. Surfaces of components of amplitude function $|2,0,½,½\rangle$; (a) first component, real; (b) third component, imaginary; (c) real part of fourth component; (d) imaginary part of fourth component; all surfaces are cut open to reveal the inner structure.

The four plots in figure 2 resemble the corresponding plots in figure 1 except that in each case there is an inner sphere, in figure 2a, or two inner hemispheres, in figures 2b,c,d; this sphere or these hemispheres are centred on the origin, which is the location of the atomic nucleus. These plots thereby resemble amplitude functions in spherical polar coordinates as follows: $\psi_{2,0,0}(r,\theta,\phi)$ that is real, $\psi_{2,1,0}(r,\theta,\phi)$ that is real, and $\psi_{2,1,1}(r,\theta,\phi)$ in its real and imaginary parts, respectively. The scalar product of $|2,0,½,½\rangle$ with its complex conjugate is

$$|2,0,\tfrac{1}{2},\tfrac{1}{2}\rangle^* \cdot |2,0,\tfrac{1}{2},\tfrac{1}{2}\rangle \;=\; \frac{1}{512}\, e^{(-r)}\,(Z^2\alpha^2 r^2 - 8 Z^2\alpha^2 r + 16 Z^2\alpha^2 + 16 r^2 - 64 r + 64)\,,$$

which consists of three terms containing $\alpha^2 Z^2$ defining the small part and three other terms that define the large part of this probability density function. We plot the large part in figure 3a, which reveals three spherical surfaces; the innermost surface defines a sphere that would hold electronic charge density, and the other two surfaces define a shell that would likewise hold electronic charge density between them. Figure 3b likewise shows three spherical surfaces for the small part divided by $\alpha^2 Z^2$ so that it is magnified to be much larger than its proportion in relation to the large part. In either case there is a surface between the innermost sphere and the outer shell on which there is zero electronic charge density, similar to a nodal plane for $\psi_{0,1,0}(r,\theta,\phi)$ at which the phase changes from positive to negative or an infinitesimally thin spherical shell of zero charge density for $\psi_{1,0,0}(r,\theta,\phi)^2$. The locations of these surfaces of zero charge density for the small and large parts of $|2,0,\tfrac{1}{2},\tfrac{1}{2}\rangle^* \cdot |2,0,\tfrac{1}{2},\tfrac{1}{2}\rangle$ differ; when these parts are combined to obtain the total squared amplitude, or the corresponding density of electronic charge, there is hence no point on a line from the nucleus from $r = 0$ to $r \to \infty$ at which there is zero amplitude squared or charge density. To illustrate this condition, figure 3c shows a greatly magnified profile of the squared amplitude along a radius $r$; despite a significantly decreased squared amplitude near $r = 2\, a_0$, no osculation of the curve and the abscissal axis occurs.

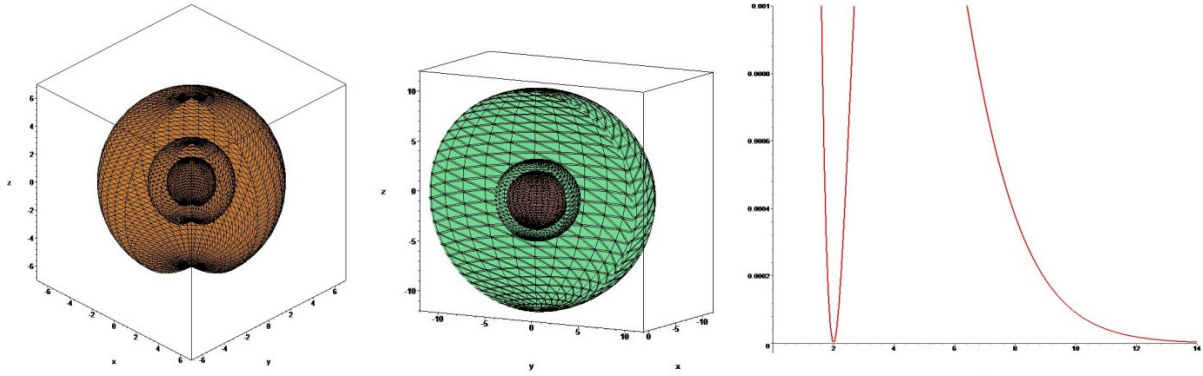

Figure 3  Surfaces of squared amplitude function $|2,0,\tfrac{1}{2},\tfrac{1}{2}\rangle$; (a) large part; (b) small part divided by $\alpha^2 Z^2$; (c) curtailed profile of total squared amplitude function $|2,0,\tfrac{1}{2},\tfrac{1}{2}\rangle$; in figures 3a and 3b, the colour change of an inner sphere is present for visual guidance but represents no physical effect.

For amplitude function $|2,1,\tfrac{1}{2},\tfrac{1}{2}\rangle$, which is represented as this vector,

$$\|2,1,\tfrac{1}{2},\tfrac{1}{2}\rangle \;=\; \begin{bmatrix} -\dfrac{1}{24}\sqrt{2}\sqrt{3}\, r\, e^{\left(-\tfrac{r}{2}\right)} \cos(\theta) \\[6pt] -\dfrac{1}{24}\sqrt{2}\sqrt{3}\, r\, e^{\left(-\tfrac{r}{2}\right)} \sin(\theta)\, e^{(i\phi)} \\[6pt] \dfrac{1}{16} i\sqrt{2}\sqrt{3}\,\alpha\, Z\left(1-\dfrac{r}{6}\right) e^{\left(-\tfrac{r}{2}\right)} \\[6pt] 0 \end{bmatrix}$$

both large components are non-zero but the fourth component is zero. The first component is purely real but the second component has real and imaginary parts; the third component, small, is purely imaginary but lacks an angular dependence. We plot these surfaces in figure 4.

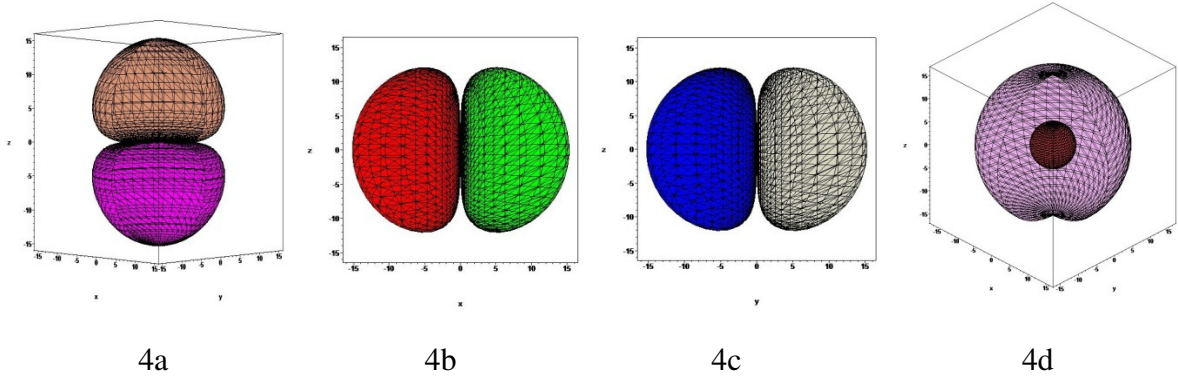

4a  4b  4c  4d

Figure 4  Surfaces of components of amplitude function $|2,1,\tfrac{1}{2},\tfrac{1}{2}\rangle$; (a) first component, real; (b) second component, real part; (c) second component, imaginary part; (d) imaginary third component; the surfaces in figures 4a and 4d are cut open to reveal the inner structure or lack thereof.

The surface of the real first component of amplitude function $|2,1,\tfrac{1}{2},\tfrac{1}{2}\rangle$ resembles $\psi_{0,1,0}(r,\theta,\phi)$; the surfaces of the second component in their real and imaginary parts that likewise lack an internal structure resemble the real and imaginary parts of $\psi_{0,1,-1}(r,\theta,\phi)$, respectively. The surface of the third and purely imaginary component of $|2,1,\tfrac{1}{2},\tfrac{1}{2}\rangle$ resembles the real function $\psi_{1,0,0}(r,\theta,\phi)$. The square of this amplitude function becomes

$$|2,1,\tfrac{1}{2},\tfrac{1}{2}\rangle^* \cdot |2,1,\tfrac{1}{2},\tfrac{1}{2}\rangle \;=\; \dfrac{1}{1536} e^{(-r)} \left( Z^2 \alpha^2 r^2 - 12\, Z^2 \alpha^2 r + 36\, Z^2 \alpha^2 + 16\, r^2 \right)$$

of which the surface of the large part in the last term of the sum plots as simply a sphere, in figure 5a, but has no exact counterpart as $\psi_{k,l,m}(r,\theta,\phi)$; because of factor $r^2$ this function should plot as a spherical shell but the inner surface of that shell is so near the origin as to be indiscernible. The surface of the small part, divided by $\alpha^2 Z^2$, plots as an inner sphere and an outer shell, with a surface of zero probability density between them, in figure 5b. The surface of the combined large and small parts of the squared amplitude function simply plots as another sphere with an indiscernible inner core.

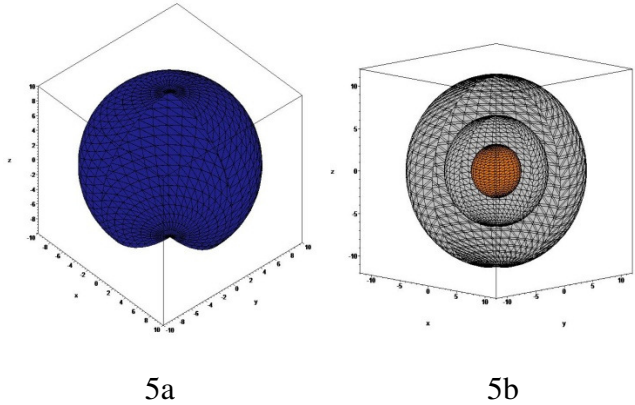

5a           5b

Figure 5  Surfaces of $|2,1,½,½>^* \cdot |2,1,½,½>$ (a) large part; (b) small part divided by $\alpha^2 Z^2$ showing an inner core and an outer shell; the colour change is intended only for visual guidance and has no physical significance.

Amplitude function $|2,1,3/2,3/2>$ has four components,

$$|2,1,3/2,3/2> = \begin{bmatrix} \frac{1}{8} r e^{\left(-\frac{r}{2}\right)} \sin(\theta) e^{(i\phi)} \\ 0 \\ \frac{1}{32} i\, \alpha\, Z\, r\, e^{\left(-\frac{r}{2}\right)} \sin(\theta) \cos(\theta) e^{(i\phi)} \\ \frac{1}{32} i\, \alpha\, Z\, r\, e^{\left(-\frac{r}{2}\right)} \sin(\theta)^2 e^{(2 i\phi)} \end{bmatrix}$$

of which all non-zero components are complex because they contain $e^{i\phi}$ or its square; the second large component is zero.  Figure 6 shows the pertinent surfaces.

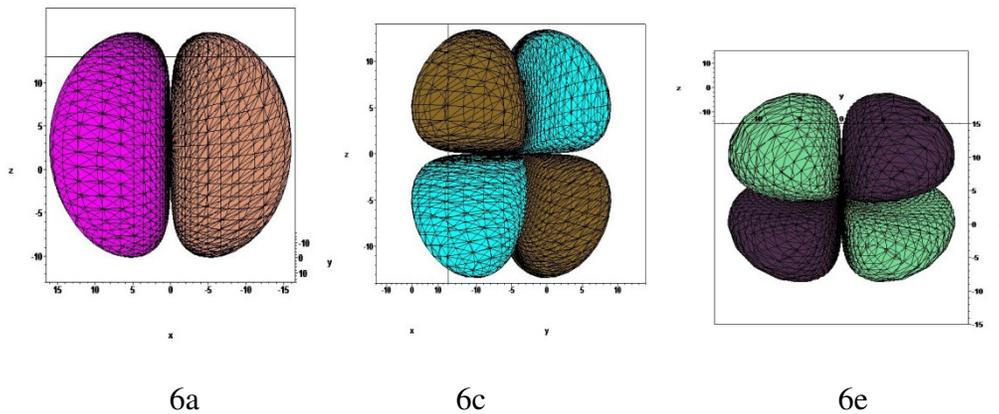

6a           6c           6e

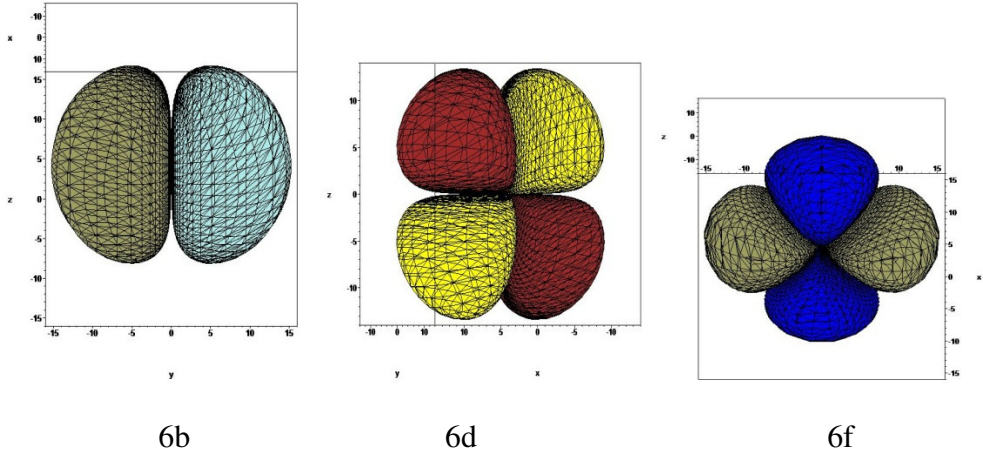

| 6b | 6d | 6f |

Figure 6 Surfaces of components of amplitude function |2,1,3/2,3/2>; (a) first component, real part; (b) first component, imaginary part; (c) third component, real part; (d) third component, imaginary component; (e) fourth component, real part; (f) fourth component, imaginary part

For |2,1,3/2,3/2> no surface contains internal structure. The surfaces of the real and imaginary parts of the first component resemble those of the real and imaginary parts of $\psi_{0,1,-1}(r,\theta,\phi)$, respectively; the surfaces of the real and imaginary parts of the third component resemble those of the real and imaginary parts of $\psi_{0,2,1}(r,\theta,\phi)$, respectively, whereas the surfaces of the real and imaginary parts of the fourth component resemble those of the real and imaginary parts of $\psi_{0,2,-2}(r,\theta,\phi)$, respectively. The scalar product of |2,1,3/2,3/2> and its complex conjugate vector yields this formula comprising only two terms,

$$|2,1,3/2,3/2>^* \cdot |2,1,3/2,3/2> = \frac{1}{1024} \sin(\theta)^2 (Z^2 \alpha^2 + 16) r^2 e^{(-r)}$$

which differ only through a scaling factor $\alpha^2 Z^2$ and a numerical coefficient; the plot of either surface, such as the large part in figure 7, shows only a torus of elliptical cross section about the polar axis with no probability density on that axis, similar to the surface of $\psi_{0,1,1}(r,\theta,\phi)^2$.

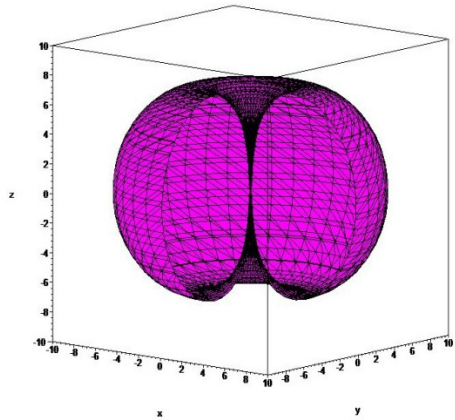

Figure 7  Surface of |2,1,3/2,3/2>* . |2,1,3/2,3/2>, large part

The other amplitude function belonging to $n = 2$ is |2,1,3/2,1/2> that has four non-zero components,

$$|2,1,3/2,1/2\rangle = \begin{bmatrix} \dfrac{1}{24}\sqrt{2}\sqrt{6}\, r\, e^{\left(-\frac{r}{2}\right)} \cos(\theta) \\ -\dfrac{1}{48}\sqrt{2}\sqrt{6}\, r\, e^{\left(-\frac{r}{2}\right)} \sin(\theta)\, e^{(i\phi)} \\ \dfrac{1}{64} i \sqrt{2}\sqrt{6}\, \alpha\, Z\, r\, e^{\left(-\frac{r}{2}\right)} \left( \cos(\theta)^2 - \dfrac{1}{2} \right) \\ \dfrac{1}{64} i \sqrt{2}\sqrt{6}\, \alpha\, Z\, r\, e^{\left(-\frac{r}{2}\right)} \sin(\theta) \cos(\theta)\, e^{(i\phi)} \end{bmatrix}$$

of which the first, large, is purely real, the third, small, is purely imaginary and the other two are complex; the surfaces are shown in figure 8.

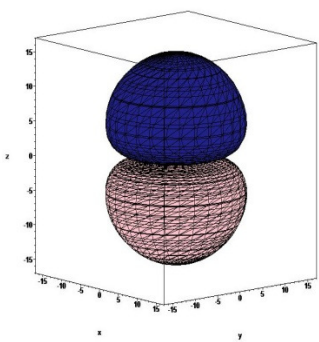 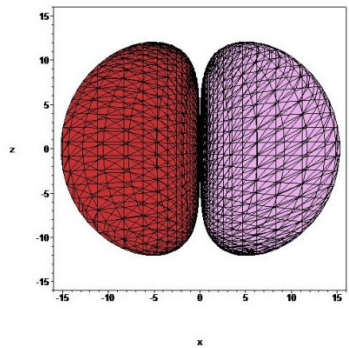 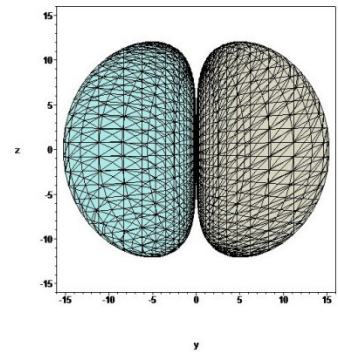

8a  8b  8c

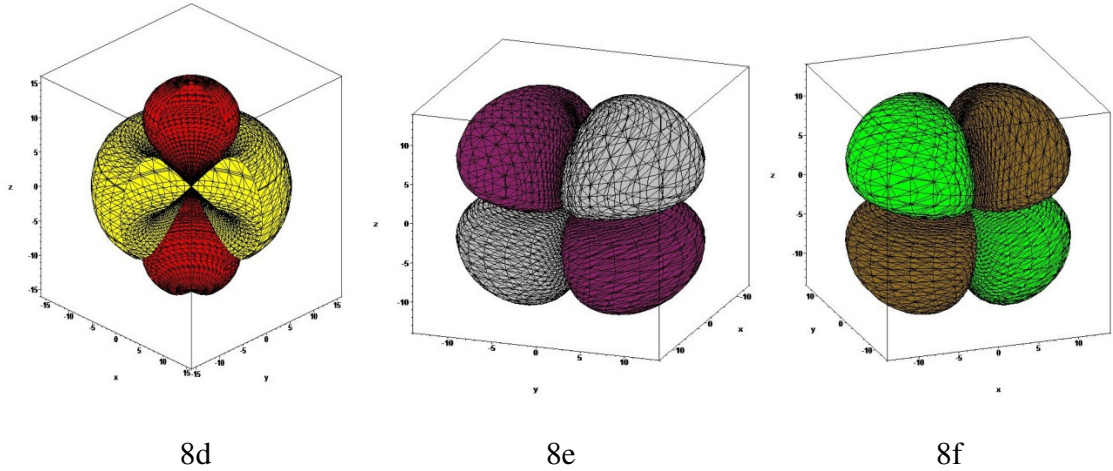

8d          8e          8f

Figure 8  Surfaces of components of amplitude function |2,1,3/2,1/2>; (a) first component, real; (b) second component, real part; (c) second component, imaginary part; (d) third component, imaginary, (e) fourth component, real part; (f) fourth component, imaginary part; the surface in 8d is cut open to reveal the internal structure.

These surfaces of components of amplitude function |2,1,3/2,1/2> resemble, in the same order, the surface of $\psi_{0,1,0}(r,\theta,\phi)$ that is real, of $\psi_{0,1,-1}(r,\theta,\phi)$ that has real and imaginary parts, of $\psi_{0,2,0}(r,\theta,\phi)$ that is real, and of $\psi_{0,2,1}(r,\theta,\phi)$ that has real and imaginary parts. The product of |2,1,3/2,1/2> and its complex conjugate,

$$|2,1,3/2,1/2> * \cdot |2,1,3/2,1/2> \; = \; \frac{1}{12288} \, r^2 \, e^{(-r)} \, (9 \, Z^2 \, \alpha^2 + 192 \, \cos(\theta)^2 + 64)$$

contains two terms for the large part and one term for the small part; the surface of that small part resembles the spherical surface of one term of the large part, which we plot in figure 9a. The surface of the other term of the large part, containing $\cos(\theta)^2$, appears in figure 9b; the surface of the sum of those two large parts appears in figure 9c.

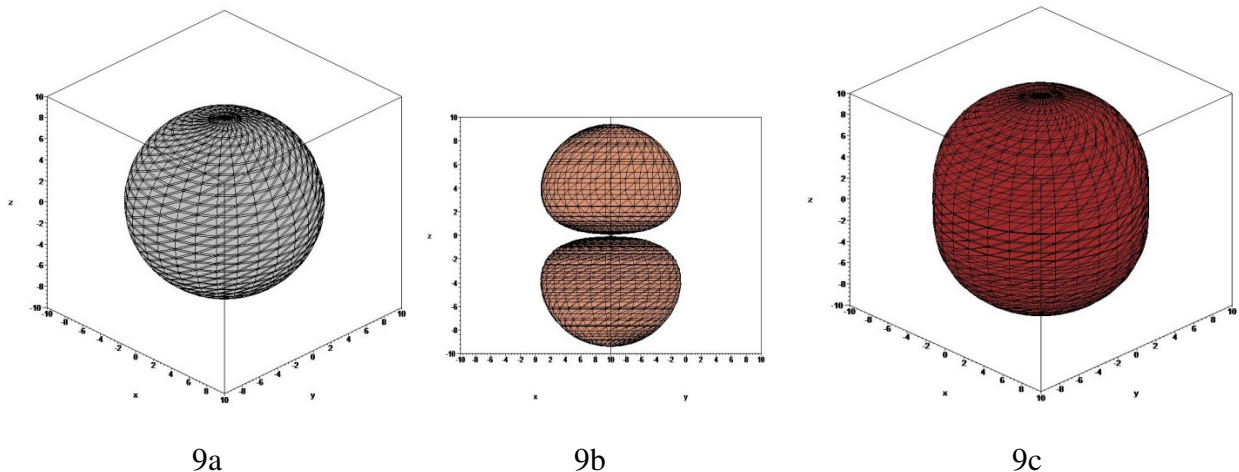

9a          9b          9c

Figure 9 Surfaces of the large part of |2,1,3/2,1/2> * . |2,1,3/2,1/2> (a) first term; (b) second term containing $\cos(\theta)^2$; (c) sum of first and second terms

The surface of that first term of squared amplitude function |2,1,3/2,1/2> resembles the surface of $\psi_{0,0,0}(r,\theta,\phi)^2$ and of the second term resembles that of $\psi_{0,1,0}(r,\theta,\phi)^2$, but the surface of the sum resembles no surface of a particular amplitude function $\psi_{k,l,n}(r,\theta,\phi)^2$.

The preceding plots of these five amplitude functions, and their squares, for the hydrogen atom with $n=1$ and $n=2$ according to relativistic wave mechanics illustrate the correlations with the amplitude functions according to non-relativistic wave mechanics.

## III  CIRCULATING CURRENT

Apart from the probability density calculated as $\psi^* \psi$ or |n,l,j,m_j>* . |n,l,j,m_j>, to which the density of electronic charge of the hydrogen atom is proportional, shown in plots of the surfaces of squared amplitude functions above, a quantum-mechanical problem implies a probability flux; in the stationary state this flux might not vanish, but its divergence must be zero so that probability is locally conserved. In units of speed of light $c$, the cartesian spatial components of this probability flux are calculated from $\psi^* \alpha_x \psi$, $\psi^* \alpha_y \psi$ and $\psi^* \alpha_z \psi$, in which velocity matrices $\alpha_x, \alpha_y, \alpha_z$ are the Dirac matrices [12, 13]. In each case, summarised in table 1 in SI notation and containing Bohr radius $a_0$ and charge $e$ on a proton with all pertinent factors, there is a finite electronic current density that is proportional to the probability flux and that circulates about the polar axis; this current density differs for each state but appears to be largest for the hydrogen atom (with $Z = 1$) in its ground electronic state. There is no current density parallel to this axis as each current density is proportional to $\sin(\theta)$ to some power. In figure 10 we plot, for each of the five states of the hydrogen atom treated above, a surface that contains 0.995 of the electronic current density, in the same way that we plot surfaces for the probability density, to which the charge density is proportional, in figures 3, 5, 7 and 9 above.

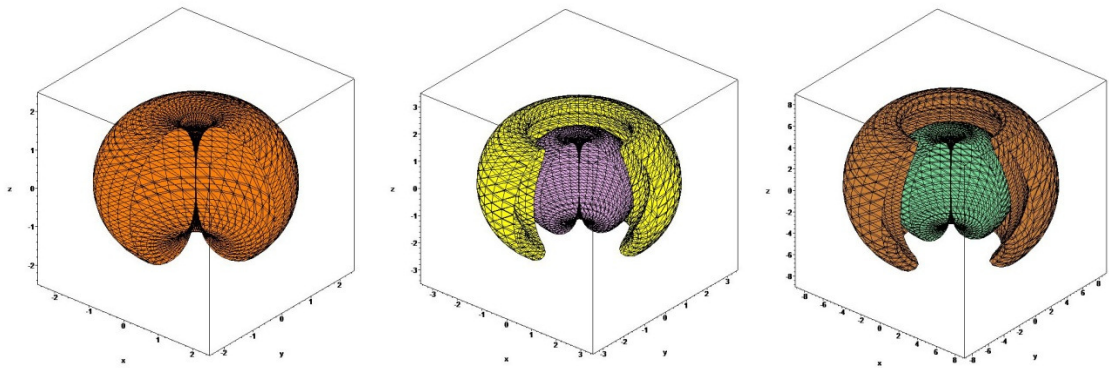

10a            10b            10c

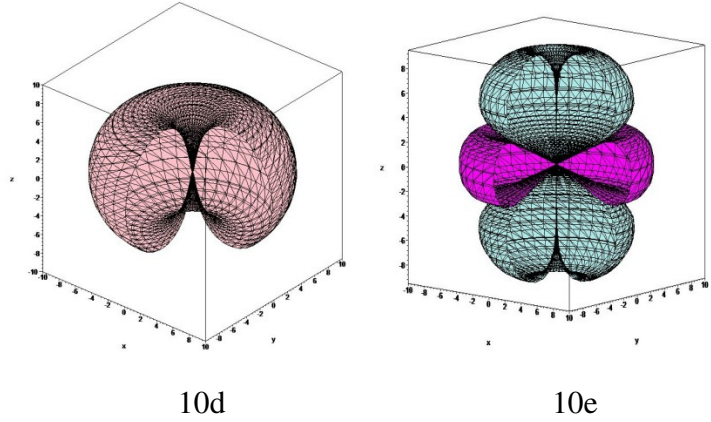

|10d|10e|

Figure 10   Surfaces of constant electronic current density for the hydrogen atom in states (a) |1,0,½,½>, (b) |2,0,½,½>, (c) |2,1,½,½>, (d) |2,1,3/2,3/2>, (e) |2,1,3/2,1/2>; in each case the surface, which is cut open to reveal the structure near the polar axis, encloses 0.995 of the total electronic current density.

The current circulates about the polar axis, $z$ in cartesian coordinates; there is no current on or along that axis. The presence of $\sin(\theta)$ in each formula for this current density ensures that the maximum current density occurs in plane $z = 0$ that contains the origin and hence the atomic nucleus that lies on that polar axis.  In each case the form of the surface is that of a torus; for state |2,0,½,½> or |2,1,½,½> there are two tori, one separate from the other; in the latter cases the current flows clockwise in one torus and anticlockwise in the other torus, with a cylindrical surface, infinitesimally thin, of zero current density between them. There is no absolute definition of the direction of either flux. These two cases correlate with the inner structure of the amplitude functions or their squares in the same two cases, for state |2,0,½,½> shown in figures 2 and 3 and for state |2,1,½,½> shown in figures 4 and 5.  The outer diameter/$a_0$ of the torus for |1,0,½,½> is approximately 4.6, comparable with the diameter of the sphere of $\psi^2$ for this state according to our standard convention of constant $\psi$. For state |2,0,½,½> or |2,1,½,½> the outer diameter of the inner torus is comparable with the diameter of the inner sphere of $\psi^2$ or the two hemispheres together, but the outer diameter of the outer torus is much smaller than the outer diameter of the surface of $\psi^2$. For state |2,1,3/2,1/2> there are three tori in a vertical stack about the polar axis, arising from factors $6\cos(\theta)^2 - 1$ and $\sin(\theta)$ in the density of circulating current, as specified in table 1; the flow of electric current hence reverses twice in a direction parallel to the polar axis. A resemblance of the torus for the electronic current density in state |2,1,3/2,3/2>, shown in figure 10d, to the probability density of the same state, shown in figure 7, is likely fortuitous.

In Schroedinger's fourth paper [3], he produced formulae for the components of the density of circulating current in spherical polar coordinates $r,\theta,\phi$ as follows in SI notation,

$$J_r = J_\theta = 0, \ J_\phi = -\frac{e\,h\,m}{2\,\pi\,\mu\,r\,\sin(\theta)}\ |\psi_{k,l,m}(r,\theta,\phi)|^2$$

such that the radial and azimuthual components are zero and the component of current flux about the polar axis is proportional to equatorial quantum number *m*. The latter property implies that for any state with *m* = 0, so including the ground state, there is no circulating current, contrary to experimental fact. This distinction between the Schroedinger and Dirac treatments of the hydrogen atom is significant; the former result is incorrect whereas the result from Dirac's theory is precisely correct. The existence of these circulating currents has an importance consequence; associated with a circulating current is a magnetic field that is characterised with a magnetic dipolar moment exactly sufficient to account for all magnetic properties of the electron in the hydrogen atom: there is no justification of an electron *spin* whereby an electron is supposed to rotate about its internal axis. Half a century ago, Powell reproached [6] the authors of then contemporary textbooks of chemistry who propagated this incorrect explanation of magnetic properties, which originated as an hypothesis in 1925 in the work of Pauli and of Goudsmit, Uhlenbeck and Kronig before the discovery of quantum mechanics, but the error regrettably persists even to this day. Further explanation of the relativistic wave mechanics and comparison with other descriptions of the hydrogen atom are available elsewhere [12].

## IV DISCUSSION

This treatment of the hydrogen atom according to Dirac's relativistic wave mechanics generates yet another set of quantum numbers -- $n, l, j, m_j$ -- and the corresponding sets of geometric shapes of surfaces of constant amplitude, beyond those of the four sets of quantum numbers and amplitude functions for the hydrogen atom according to non-relativistic wave mechanics in the four respective systems of coordinates with separable variables [7], not to mention a solution with incompletely separated spatial variables [13, 14]. Even when Dirac's equation is solved in systems of coordinates other than spherical polar, these four quantum numbers remain pertinent because they specify a state of the hydrogen atom, independently of any other quantum numbers that might pertain to particular amplitude functions. This diversity of shapes of the surface of an amplitude function or *orbital* makes clear that such a shape, even for a specific state characterised with a particular energy and angular momentum, has no absolute meaning; any meaning is confined to a specific context according to a selected system of coordinates and method of quantum mechanics. Erroneous ideas that originated as hypotheses before experimental or theoretical tests were practicable have persisted, largely because of the superficial scholarship of succeeding generations of scientists and authors, and their inadequate mathematical understanding of primarily mathematical theories and properties. To alleviate this deprecated condition one must encourage the use of advanced mathematical software; the latter means provides a chemist or physicist, who is not primarily a mathematician but who has a sufficient understanding of mathematical concepts and principles, with a strong capability to undertake the appropriate mathematical operations and to derive correct and exact results that can be made subject to experimental and theoretical test. The hydrogen atom is a simple system for experimental or theoretical investigation; such simplicity should not tempt one to infer properties beyond the defined limits of that system.

## ACKNOWLEDGEMENT

I thank Professor J. D. Hey for helpful discussion.

**Table 1** Magnitude of circulating electronic current for the hydrogen atom ($Z=1$) in five states

| state or amplitude function | magnitude of circulating current |
|---|---|
| $\|1,0,1/2,\pm 1/2\rangle$ | $\dfrac{e\,c\,\alpha\,Z^4 \sin(\theta)\,e^{\left(-\frac{2Zr}{a_0}\right)}}{\pi\,a_0^{\,3}}$ |
| $\|2,0,1/2,\pm 1/2\rangle$ | $\dfrac{e\,c\,\alpha\,Z^4 \left(-4+\dfrac{Zr}{a_0}\right)\left(\dfrac{Zr}{a_0}-2\right)\sin(\theta)\,e^{\left(-\frac{Zr}{a_0}\right)}}{64\,\pi\,a_0^{\,3}}$ |
| $\|2,1,1/2,\pm 1/2\rangle$ | $\dfrac{e\,c\,\alpha\,Z^5 \left(-6+\dfrac{Zr}{a_0}\right) r \sin(\theta)\,e^{\left(-\frac{Zr}{a_0}\right)}}{192\,\pi\,a_0^{\,4}}$ |
| $\|2,1,3/2,\pm 3/2\rangle$ | $\dfrac{e\,c\,\alpha\,Z^6\,r^2\,e^{\left(-\frac{Zr}{a_0}\right)}\sin(\theta)^3}{128\,\pi\,a_0^{\,5}}$ |
| $\|2,1,3/2,\pm 1/2\rangle$ | $\dfrac{e\,c\,\alpha\,Z^6\,r^2 \sin(\theta)\,(6\cos(\theta)^2 - 1)\,e^{\left(-\frac{Zr}{a_0}\right)}}{256\,\pi\,a_0^{\,5}}$ |